# Locally Lorentz-Covariant Theory of Gravity Founded on Inertial Frame of Center of Mass

Hai-Long Zhao

**Abstract:** A locally Lorentz-covariant theory of gravity that is equivalent to general relativity in weak gravitational field is suggested. The space-time standards in local gravitational field are modified in terms of equivalence principle to keep them consistent with those of inertial frame. The static metric in our theory agrees with Schwarzschild metric to the first order approximation. According to our metric expression, black hole and singularity do not exist. The gravitational vector potential generated by a moving body is obtained by applying local Lorentz transformation to Schwarzschild metric in rectangular coordinate system. In our theory, the center of mass of the system is taken as the inertial reference frame. When observed from center of mass, the results of periastron precession and gravitational radiation of binary star system are different from those of general relativity, which are derived from the relative motion of the binary. What's more, the expansion of the universe should also be observed from the center of the universe. By assuming that the Hubble constant varies with different evolution stage of the universe, dark energy is not needed.

## 1 Introduction

Einstein's general relativity is a generally covariant theory of gravity. While successfully explaining some phenomena that cannot be explained in Newtonian theory, it has also some unsatisfying points: The first is the interpretation of gravity. Gravity is taken as the consequence of curvature of space-time in general relativity, which is different from the other three forces: electromagnetic force, strong force and weak force. Second, there exist black hole and singularity in general relativity. Once an object falls into a black hole, the gravitational force it experiences is infinite, it will inevitably fall to the center, forming a singularity with infinitesimal volume and infinite density. The existence of singularity is a deep-rooted problem in general relativity, which cannot be eliminated by coordinate transformation or other methods, and it may also violate the uncertainty principle. Third, the mathematical formalism of general relativity is too complex for most people.

Then where to find a simpler theory of gravity? We might start with the combination of the law of universal gravitation and the special theory of relativity. Historically, someone had made unsuccessful attempts in this way, but they did not consider the influence of gravitational field on space-time; they only considered the relativistic effect. In special relativity, there are time dilation and length contraction. In the gravitational field, the same effects also exist. Einstein first noticed this when he established the general theory of relativity and pointed out that there are no synchronous clocks and unified rulers in the gravitational field. If we combine the relativistic effect and gravity effect with the law of universal gravitation, we should obtain the same results as those of general relativity. This is the idea of introducing a new theory of gravity.

Email: zhlzyj@126.com

## 2 Static Gravitational Field

In classical electrodynamics, photons are uncharged, the existence of electromagnetic field does not affect light. Therefore, synchronous clocks and unified rulers can be established everywhere in space, and Lorentz transformation can be applied to large distances. While in the gravitational field, light will also be affected by gravity. In order to see the effect of gravity on light, let's start with the more general situation of the effect of gravity on massive particles.

### 2.1 Static gravitational potential

We know that Coulomb's law applies to the interaction between rest charges, but can be extended to the interaction of a rest charge on a moving charge. Similarly, we suppose Newton's law of universal gravitation also holds for the interaction of a rest gravitational source on a moving body. We suppose, for simplicity, that the center of a spherically symmetric body with the mass of $M$ is fixed at point O, as shown in Fig. 1, and a test body with the gravitational mass of $m$ is falling freely in the static gravitational field generated by $M$. As it falls, the work done on it by the gravitational field will be converted into its kinetic energy, and its inertial mass increases. Then according to the principle of equivalence, its gravitational mass also increases and it will experience a larger gravitational force.

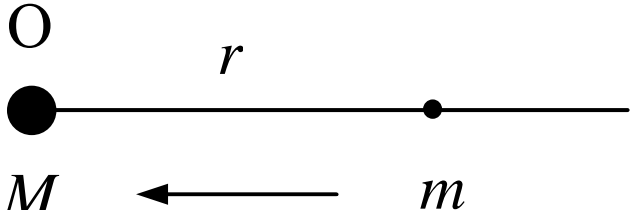

Fig. 1 A test body falls freely in the static gravitational field.

Suppose the body $m$ moves a differential distance of $\mathrm{d}r$, the change of mass is $\mathrm{d}m$ and the change of kinetic energy is $\mathrm{d}mc^2$. The work done by the gravitational field is $-GMm\mathrm{d}r/r^2$. According to the relationship between work and energy, we have

$$\mathrm{d}mc^2 = -\frac{GMm}{r^2}\mathrm{d}r \tag{1}$$

which follows that

$$\frac{\mathrm{d}m}{m} = -\frac{GM}{r^2c^2}\mathrm{d}r \tag{2}$$

Integrating on both sides of the equation, we find

$$m = m_0 \mathrm{e}^{\frac{GM}{rc^2}} \tag{3}$$

where $m_0$ is the rest mass at infinity.

The reason we emphasize static gravitational field is that we regard $M$ as a rest inertial frame and $m$ a test particle. In fact, $M$ will also be attracted by $m$. So $M$ is actually not an inertial frame.

Eq. (3) holds only for the situation of $M >> m$. If $m$ and $M$ are comparable, we must observe the motions of the two bodies from the center of mass of the system, which is an inertial reference frame. We shall discuss this issue later.

The effective mass of a photon is $\hbar\omega/c^2$. Suppose the frequency of light at infinity is $\omega_0$. As it travels in the gravitational field, its frequency will be $\omega = \omega_0 e^{GM/rc^2}$ according to Eq. (3). Likewise, if the frequency of light at the surface of a star with the radius $R$ is $\omega_0$, it will become $\omega = \omega_0 e^{-GM/Rc^2} \approx \omega_0(1 - GM/Rc^2)$ as it propagates to infinity, which is the formula of gravitational redshift.

The expression of gravitational potential is $-GM/r$ in Newtonian mechanics. Now we recalculate it. By definition, the potential energy of a body equals to the negative value of the work done by the gravitational field as it falls from infinity. Then we have

$$E_p = -\int_{\infty}^{r} -\frac{GMm}{r^2} dr = \int_{\infty}^{r} \frac{GMm_0}{r^2} e^{\frac{GM}{rc^2}} dr = m_0 c^2 (1 - e^{\frac{GM}{rc^2}}) \tag{4}$$

Thus the static gravitational potential is $c^2(1 - e^{GM/rc^2})$, which agrees with Newtonian gravitational potential to the first order approximation.

### 2.2 The time dilation and length contraction in static gravitational field

As we have shown above, light will be affected as it travels in the gravitational field, so we cannot define synchronous clocks and unified rulers as in the inertial frames. As pointed out in [1], the time dilation effect cannot be observed by merely measuring the time interval of the ticks of a clock and comparing it with the time standard defined by the maker, because the gravitational influence on the time standard is equal to that on the clock. That is to say, if a clock reads one second for a physical process without gravity, it will still read one second in the presence of gravitational field, since the gravitational field has the same influence on the clock and the process.

In order to synchronize the clocks in the gravitational field, it is necessary to take one clock as the standard, and adjust other clocks to synchronize with it. It is better to use the clock at infinity as the time standard, because the gravity at infinity is zero, and here is an inertial frame. We calibrate the clocks at rest in the gravitational field with the clock at infinity, thus the time interval each clock measures would be the proper time that the physical process takes place in the inertial frame. Then we may apply local Lorentz transformation in the gravitational field, as will be discussed later.

Suppose there is a light wave with the frequency of $\omega_0$ at infinity, and the time interval between two adjacent wave crests is $\Delta t = 2\pi/\omega_0$. Let the light wave propagates from infinity to position $r$. Since the intervals required for the two wave crests to travel from infinity to $r$ are equal, the time interval between the two wave crest measured by the local clock at $r$ is also $\Delta t$, which shows that the gravitational field has the same influence on the process and the clock. But according to the above calculation, when observed from point O, the frequency of the light wave becomes $\omega_0 e^{GM/rc^2}$. As point O is at rest relative to infinity, and the gravitational force at O vanishes (the gravity at the center of a spherically symmetric body is zero), point O is a still inertial frame, and the clock at O is consistent with the clock at infinity. Therefore, when observed from infinity, the frequency of the light wave is $\omega_0 e^{GM/rc^2}$ and the interval between two local wave crests is $e^{-GM/rc^2}\Delta t$. In order to be consistent with the clock at infinity, the time interval measured by the local clock (which is $\Delta t$) should be multiplied by a factor of $e^{-GM/rc^2}$.

We then turn to the length contraction effect. Suppose there are a standard ruler and a pole. When there is no gravitational field, the length of the pole measured by the ruler is one meter. Then we place the pole and the ruler along the radial direction of the gravitational field, the length of the pole measured by the ruler is still one meter, i.e. the ruler and the pole experience the same gravitational contraction. In order to see the length contraction effect, we compare the distance that light travels in local gravitational field with that light travels at infinity during a same time interval. Suppose the wavelength of light at infinity is $\lambda_0$, and the distance it travels in unit time is $n\lambda_0$. When it propagates to the position of $r$, the number of the wave passing through a fixed point in unit time is still $n$. However, when observed from inertial frame at infinity, the local frequency changes, and the local wavelength becomes $e^{-GM/rc^2}\lambda_0$, that is, when observed from the distant inertial frame, the distance of the light propagating in the gravitational field in unit time is $e^{-GM/rc^2} n\lambda_0$. Length is defined as the distance that light travels in a certain period of time. Then compared with the length at infinity, the length in the gravitational field becomes smaller. In order to be consistent with the proper length at inertial frame (which is $n\lambda_0$), we should multiply the length along the radial direction of the gravitational field (which is $e^{-GM/rc^2} n\lambda_0$) by a factor of $e^{GM/rc^2}$. Since light is not affected when it propagates along the transverse direction of the gravitational field, the transverse length does not contract.

To summarize, in order to establish unified time and length standards in the gravitational field, we must multiply the time measured by the local clock by a factor of $e^{-GM/rc^2}$, and the length measured by the local ruler by a factor of $e^{GM/rc^2}$ along the radical direction. In this way, the time and length measured by the rest clock and ruler in the gravitational field would be consistent with the proper time and proper length in the inertial frame, respectively.

### 2.3 The motion equation in static gravitational field

In special relativity, the expression of infinitesimal four-dimensional interval (line element) is

$$\mathrm{d}s^2 = c^2\mathrm{d}t^2 - \mathrm{d}x^2 - \mathrm{d}y^2 - \mathrm{d}z^2 \tag{5}$$

which is a scalar. For central force, it is more convenient to use the spherical coordinates, and the above expression can be written as

$$\mathrm{d}s^2 = c^2\mathrm{d}t^2 - \mathrm{d}r^2 - r^2\mathrm{d}\theta^2 - r^2\sin^2\theta\mathrm{d}\varphi^2 \tag{6}$$

After taking into account time dilation and length contraction in the gravitational field, it should be modified as

$$\mathrm{d}s^2 = c^2 e^{-\frac{2GM}{rc^2}}\mathrm{d}t^2 - e^{\frac{2GM}{rc^2}}\mathrm{d}r^2 - r^2\mathrm{d}\theta^2 - r^2\sin^2\theta\mathrm{d}\varphi^2 \tag{7}$$

Then it is still a scalar, and it will become Schwarzschild metric to the first order approximation:

$$\mathrm{d}s^2 = c^2(1-\frac{2GM}{rc^2})\mathrm{d}t^2 - (1-\frac{2GM}{rc^2})^{-1}\mathrm{d}r^2 - r^2\mathrm{d}\theta^2 - r^2\sin^2\theta\mathrm{d}\varphi^2 \tag{8}$$

Now let's find the equation of motion for a free falling body in the static gravitational field. In the inertial frame, the free particle moves at a uniform speed along a straight line, and its equation of motion is the geodesic equation in four-dimensional space-time, which can be derived from the variational principle

$$\delta\int \mathrm{d}s = 0 \tag{9}$$

Since a freely falling body is a local inertial system, the above equation holds also for the motion of a body falling freely in the gravitational field. This generalization can be regarded as a postulate in general relativity [2]. For the motion of light, we also have another equation

$$\mathrm{d}s = 0 \tag{10}$$

The above two equations are the equations of motion for free falling bodies (including light) in the static gravitational field. At present, the test of general relativity mainly includes the following experiments:

(i) Gravitational redshift
(ii) Precession of planetary perihelion
(iii) Gravitational deflection of light
(iv) Delay of radar echo
(v) Precession of gyroscope installed on the satellite
(vi) Gravitational wave detection

Of these experiments, gravitational redshift mainly tests the equivalence principle, which has little to do with general relativity. The second to fourth experiments test the static gravitational field, while the last two test the dynamic gravitational field. Since Eqs. (7) and (8) are the same for weak gravitational field, the results of the second to fourth experiments are the same for the two theories without further calculation. For those who are interested in the calculation process, a simpler and easier method is given to calculate the perihelion of the planets and the deflection of light in the following. The same method may be used to calculate the delay of radar echo.

### 2.4 Precession of planetary perihelion

The revolution of the planet can be regarded as the movement of a point particle. At first glance, it seems that there is no precession. As the orbit of the planet is an ellipse, the orientation of the major axis is a parameter of the planetary motion. The observations show that the orientation of the major axis also rotates in space, which is the precession of the planet. In order to better understand the precession of the planet, one may reach out left (or right) hand and imagine that the planet is moving along the edge of the palm, and the middle finger represents the major axis of the ellipse. Then slowly rotate the palm, which is equivalent to the precession of the planet. This precession angular velocity can be calculated by solving Eq. (9), which has been introduced in many textbooks. Now we use a more simple and easy method to calculate the precession of the planet.

We first write the equation of conservation of energy and angular momentum of the planet in the absence of gravity:

$$\frac{1}{2}m[(\frac{\mathrm{d}r}{\mathrm{d}t})^2+(\frac{r\mathrm{d}\varphi}{\mathrm{d}t})^2]-\frac{GMm}{r}=E_0 \tag{11}$$

$$mr^2\frac{\mathrm{d}\varphi}{\mathrm{d}t}=L_0 \tag{12}$$

Then we take the gravity effect into account. Making the following substitutions: $r\to re^{GM/rc^2}$, $\mathrm{d}r\to \mathrm{d}r\mathrm{e}^{GM/rc^2}$, $\mathrm{d}t\to\mathrm{d}t\mathrm{e}^{-GM/rc^2}$, $r\mathrm{d}\varphi\to r\mathrm{d}\varphi$, $E_0\to E$, $L_0\to L$, $-\frac{GMm}{r}\to mc^2(1-\mathrm{e}^{GM/rc^2})$, we obtain

$$\frac{1}{2}m[\mathrm{e}^{\frac{4GM}{rc^2}}(\frac{\mathrm{d}r}{\mathrm{d}t})^2+\mathrm{e}^{\frac{2GM}{rc^2}}(\frac{r\mathrm{d}\varphi}{\mathrm{d}t})^2]+mc^2(1-\mathrm{e}^{\frac{GM}{rc^2}})=E \tag{13}$$

$$mr^2\mathrm{e}^{\frac{2GM}{rc^2}}\frac{\mathrm{d}\varphi}{\mathrm{d}t}=L \tag{14}$$

It should be noted that for a free falling body, the change of the radial mass has been converted into the modification of the space-time according to Eq. (1), while the transverse mass is not affected by the gravitational field, so $m$ in the above two equations is the proper mass $m_0$ but not the relativistic mass $m_0/\sqrt{1-\beta^2}$. It follows $\frac{\mathrm{d}\varphi}{\mathrm{d}t}=\frac{L}{mr^2}e^{\frac{-2GM}{rc^2}}$ from Eq. (14), and $\frac{\mathrm{d}r}{\mathrm{d}t}$ can be written as $\frac{\mathrm{d}r}{\mathrm{d}t}=\frac{\mathrm{d}r}{\mathrm{d}\varphi}\frac{\mathrm{d}\varphi}{\mathrm{d}t}$. Substituting the two expressions into (13) yields

$$(\frac{\mathrm{d}r}{\mathrm{d}\varphi})^2\frac{L^2}{m^2r^4}+r^2\mathrm{e}^{-\frac{2GM}{rc^2}}\frac{L^2}{m^2r^4}+mc^2(1-\mathrm{e}^{\frac{GM}{rc^2}})=\frac{2E}{m} \tag{15}$$

As $\frac{GM}{rc^2}<<1$, it's enough to expand $e^{-\frac{2GM}{rc^2}}$ and $e^{\frac{GM}{rc^2}}$ to the first order. Let $u=\frac{1}{r}$, we obtain

$$(\frac{\mathrm{d}u}{\mathrm{d}\varphi})^2+u^2-\frac{2GM}{c^2}u^3-\frac{2GMm^2}{L^2}u=\frac{2Em}{L^2} \tag{16}$$

Differentiating the above expression with respect to $\varphi$ yields

$$2(\frac{\mathrm{d}^2u}{\mathrm{d}\varphi^2})\frac{\mathrm{d}u}{\mathrm{d}\varphi}+2u\frac{\mathrm{d}u}{\mathrm{d}\varphi}-6u^2\frac{GM}{c^2}\frac{\mathrm{d}u}{\mathrm{d}\varphi}-\frac{2GMm^2}{L^2}\frac{\mathrm{d}u}{\mathrm{d}\varphi}=0 \tag{17}$$

Eliminating $2\frac{\mathrm{d}u}{\mathrm{d}\varphi}$, we get

$$\frac{d^2 u}{d\varphi^2} + u - \frac{3GM}{c^2}u^2 - \frac{GMm^2}{L^2} = 0 \tag{18}$$

This equation is the same as that in general relativity, so we may follow the method in general relativity. To the first order approximation, let $u = \frac{GMm^2}{L^2}$. For higher-order approximation, we assume $u = \frac{GMm^2}{L^2} + u_1$. Substituting it into the above equation yields

$$\frac{d^2 u_1}{d\varphi^2} + \frac{GMm^2}{L^2} + u_1 - \frac{3GM}{c^2}(\frac{GMm^2}{L^2} + u_1)^2 - \frac{GMm^2}{L^2} = 0 \tag{19}$$

Ignoring the higher-order term containing $u_1^2$, we obtain

$$\frac{d^2 u_1}{d\varphi^2} + (1 - \frac{6G^2M^2m^2}{c^2L^2})u_1 = 0 \tag{20}$$

The solution is $u_1 = k\cos\sqrt{1 - \frac{6G^2M^2m^2}{c^2L^2}}\varphi$, where $k$ is a constant. Thus the solution of Eq. (18) is

$$u = \frac{GMm^2}{L^2} + k\cos\sqrt{1 - \frac{6G^2M^2m^2}{c^2L^2}}\varphi \tag{21}$$

It can be seen that after $\varphi$ rotates $2\pi$, the planet must rotate forward another angle $\Delta$ to return to its initial position at $\varphi = 0$. Then we have

$$\sqrt{1 - \frac{6G^2M^2m^2}{c^2L^2}}(2\pi + \Delta) = 2\pi \tag{22}$$

As $\sqrt{1 - \frac{6G^2M^2m^2}{c^2L^2}} \approx 1 - \frac{3G^2M^2m^2}{c^2L^2}$, we get

$$\Delta = \frac{6\pi G^2M^2m^2}{c^2L^2} \tag{23}$$

Let's find the expression of angular momentum $L$. Since $L$ is conserved, for the convenience of calculation, we may calculate it at the perihelion or aphelion. Then we have

$$\frac{mv^2}{R} = \frac{GMm}{r^2} \tag{24}$$

where $R$ is the radius of curvature at the perihelion or aphelion. The equation of the ellipse can be written as

$$\frac{x^2}{a^2}+\frac{y^2}{b^2}=1 \tag{25}$$

where $a$ and $b$ are the semi-major axis and semi-minor axis of the ellipse, respectively. The expression of radius of curvature is

$$R=\frac{(1+y')^2}{|y''|} \tag{26}$$

So we have $R=b^2/a=a(1-e^2)$ at both perihelion and aphelion, where $e$ is eccentricity of the ellipse. The angular momentum of the planet at perihelion or aphelion is

$$L=mvr=m\sqrt{GMR}=m\sqrt{GMa(1-e^2)} \tag{27}$$

Substituting it into Eq. (23) yields

$$\Delta=\frac{6\pi GM}{c^2a(1-e^2)} \tag{28}$$

According to Kepler's third law, $a^3/T^2=GM/4\pi^2$ where $T$ is the revolution period of the planet, we obtain $GM=4\pi^2a^3/T^2$. Substituting it into the above formula, we get

$$\Delta=\frac{24\pi^3a^2}{c^2T^2(1-e^2)} \tag{29}$$

which is the angular shift of the perihelion after one revolution. This value is very small, but precession has a cumulative effect. After a long time, e.g. one hundred years, we may find its obvious change. The precession of Mercury is the largest in the solar system, and the theoretical value is $43''/100$ years, which agrees well with the observed one. The observed values of other planets are also consistent with the theoretical ones. For example, the theoretical value of the earth's precession is $3.84''/100$ years, while the observed value is $5.0''\pm1.2''$.

### 2.5 Deflection of light

The polar coordinate system is established with the sun as the pole and the point closest to the sun as the polar axis, as shown in Fig. 2. It can be seen that $\alpha$ is the desired deflection angle. We know that the speed of light in the inertial frame is $c$, i.e $(\mathrm{d}r/\mathrm{d}t)^2+(r\mathrm{d}\varphi/\mathrm{d}t)^2=c^2$. Considering the time dilation and length contraction in the gravitational field, it should be modified as

$$e^{\frac{4GM}{rc^2}}(\frac{\mathrm{d}r}{\mathrm{d}t})^2+e^{\frac{2GM}{rc^2}}(\frac{r\mathrm{d}\varphi}{\mathrm{d}t})^2=c^2 \tag{30}$$

In combination with Eq. (14) we obtain

$$\frac{\mathrm{d}^2u}{\mathrm{d}\varphi^2}+u-\frac{3GM}{c^2}u^2=0 \tag{31}$$

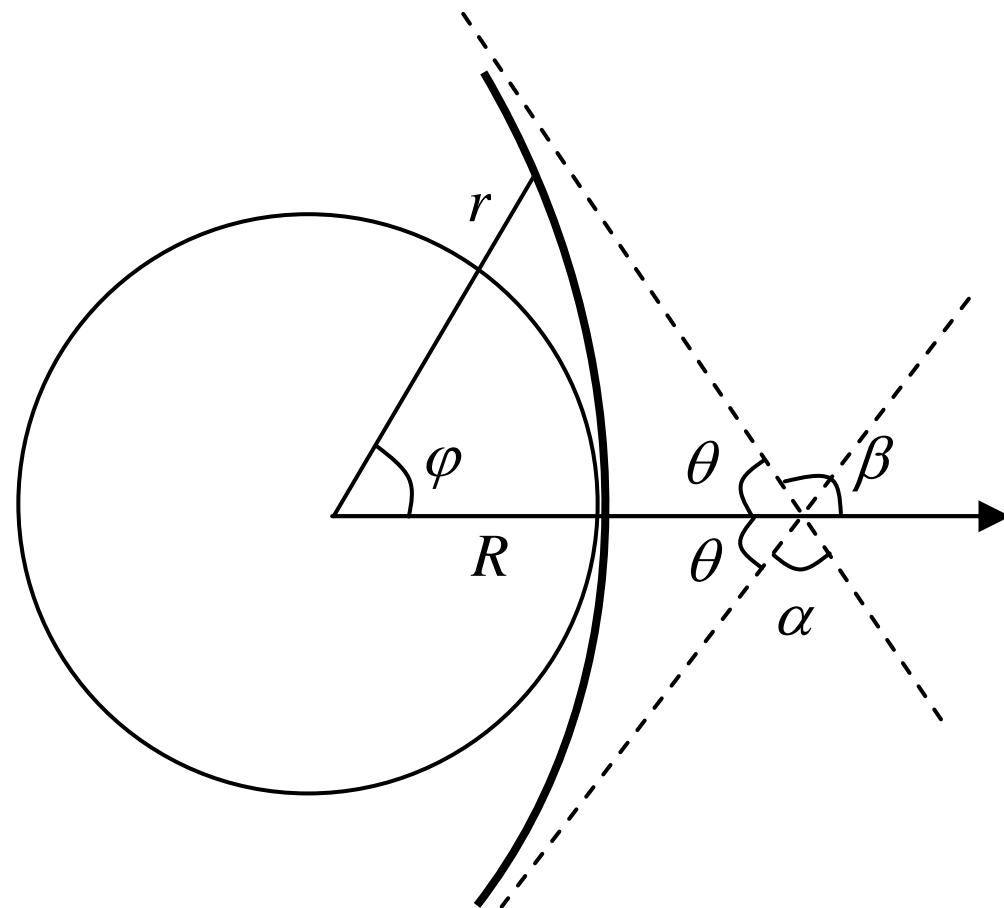

Fig. 2 The gravitational deflection of light.

which is the same as that in general relativity, so we may follow the method in general relativity. If we ignore the higher-order term containing $u^2$, the solution of the equation is $u = k\cos\varphi$, which represents a straight line, so the role of the higher-order term is to make the light deviate from the straight line. To the first order approximation we let $u_0 = R^{-1}\cos\varphi$, where $R$ is the radius of the sun. For the higher-order approximation, let $u = u_0 + u_1$. Substituting it into Eq. (31), and ignoring the higher-order term, we have

$$\frac{\mathrm{d}^2 u_1}{\mathrm{d}\varphi^2} + u_1 - \frac{3GM}{c^2 R^2}\cos^2\varphi = 0 \tag{32}$$

Let the formal solution be $u_1 = A\sin^2\varphi + B$. Substituting it into the above equation, we obtain $A = B = GM / c^2 R^2$, then we have

$$u = \frac{1}{R}\cos\varphi + \frac{GM}{c^2 R^2}(\sin^2\varphi + 1) \tag{33}$$

In the case of $r = \infty$, or $u = 1/r = 0$, the above equation corresponds to the light coming from infinity, we have

$$0 = \frac{1}{R}\cos\varphi + \frac{GM}{c^2 R^2}(\sin^2\varphi + 1) \tag{34}$$

As the light comes from infinity, $\varphi$ is very close to $\pi/2$. Let $\varphi = \pi/2 + \Delta$, since $\Delta$ is very small, we have

$$\cos\varphi = \cos(\frac{\pi}{2} + \Delta) = \cos\frac{\pi}{2}\cos\Delta - \sin\frac{\pi}{2}\sin\Delta = -\sin\Delta \approx -\Delta \tag{35}$$

$$\sin\varphi = \sin(\frac{\pi}{2} + \Delta) = \sin\frac{\pi}{2}\cos\Delta + \cos\frac{\pi}{2}\sin\Delta = \cos\Delta \approx 1 \tag{36}$$

Substituting them into Eq. (34), we have

$$0=-\frac{\Delta}{R}+\frac{2GM}{c^2R^2} \tag{37}$$

which follows that $\Delta=\frac{2GM}{Rc^2}$. As can be seen from Fig. 2, there are two asymptotes of light, one is the incoming ray and the other is the outgoing ray. The included angle between the two asymptotes is just the deflection angle of light. $\beta$ is the angle between one of the asymptotes and the polar axis. Since $\varphi$ and $\beta$ correspond to the light coming from the same point at infinity, the two rays are approximately parallel, so we have $\beta\approx\varphi=\frac{\pi}{2}+\Delta$. Now let's find the relationship between $\alpha$ and $\beta$. It can be seen from Fig. 2 that $\beta+\theta=\pi$, $\alpha+2\theta=\pi$, so we have $\alpha=2\beta-\pi=2\Delta=\frac{4GM}{Rc^2}$. Substituting the mass and radius of the sun into the formula, we obtain the deflection angle $\alpha=1.75''$.

**2.6 Precession of binary**

In the solar system, the mass of the sun accounts for 99% of the total mass, the influence of planets on the sun can be ignored, the sun may be regarded as an inertial frame. For the binary star system, the masses of the two component stars are comparable, both stars rotate around their common center of mass. There is a significant difference between the precession of the binary star system and that of the planet, which will be discussed in detail below.

For the binary star system, in addition to the gravity effect that can cause the precession of the periastron, the tidal motion and spin of the star itself will also cause the precession of the periastron, which has been analyzed and calculated in detail in Ref. [3]. We directly follow the method in [3]. The relative motion equation of the binary is:

$$\ddot{r}-r\dot{\varphi}^2=-\frac{G(m_1+m_2)}{r^2}(1+\delta r^{-2}+\delta' r^{-5}) \tag{38}$$

$$r^2\dot{\varphi}=h \tag{39}$$

where $\delta=\frac{k_1R_1^5}{Gm_1}\dot{\theta}_1^2+\frac{k_2R_2^5}{Gm_2}\dot{\theta}_2^2$, $\delta'=\frac{6m_2}{m_1}k_1R_1^5+\frac{6m_1}{m_2}k_2R_2^5$, $k_1$ and $k_2$ are constants, $R_1$ and $R_2$ are respectively the radii of the two components, $m_1$ and $m_2$ are their masses, $\dot{\theta}_1$ and $\dot{\theta}_2$ are the spin angular velocities. Let $u=1/r$, we obtain

$$\frac{\mathrm{d}^2u}{\mathrm{d}\varphi^2}+u=\frac{G(m_1+m_2)}{h^2}(1+\delta u^2+\delta' u^5) \tag{40}$$

For each revolution of the binary, the angle shift of the periastron is:

$$2\pi\varepsilon\approx 2\pi[\delta l^{-2}+\frac{5}{2}\delta' l^{-5}(1+\frac{3}{2}e^2+\frac{1}{8}e^4)] \tag{41}$$

where $l=a(1-e^2)$, $a$ is semi-major axis of the relative motion orbit, $e$ is eccentricity. Now we observe the motion of the binary from the center of mass of the system. First, let's find the equation of motion of $m_1$ relative to the center of mass. In terms of $m_1 r_1 = m_2 r_2$ and $r_1 + r_2 = r$, we get $r = (1 + m_1 / m_2) r_1$. Substituting it into Eq. (38), we obtain the equation of $m_1$ relative to the center of mass. Similarly, we can write the conservation equation of angular momentum of $m_1$ with respect to the center of mass, then we have

$$\begin{cases} \ddot{r}_1 - r_1 \dot{\varphi}_1{}^2 = -\dfrac{G m_2^3}{(m_1 + m_2)^2 r_1{}^2}(1 + \delta_1 r_1{}^{-2} + \delta_1' r_1{}^{-5}) \\ r_1{}^2 \dot{\varphi}_1 = h_1 \end{cases} \tag{42}$$

where $\dot{\varphi}_1 = \dot{\varphi}$, $\delta_1 = (\dfrac{m_2}{m_1 + m_2})^2 \delta$, $\delta_1' = (\dfrac{m_2}{m_1 + m_2})^5 \delta'$, $h_1 = (\dfrac{m_2}{m_1 + m_2})^2 h$. Let $u_1 = 1 / r_1$, we obtain

$$\frac{\mathrm{d}^2 u_1}{\mathrm{d}\varphi_1^2} + u_1 = \frac{G m_2^3}{(m_1 + m_2)^2 h_1^2}(1 + \delta_1 u_1^2 + \delta_1' u_1^5) = \frac{G m_2}{h^2}(1 + \delta_1 u_1^2 + \delta_1' u_1^5) \tag{43}$$

Comparing Eq. (40) to (43), and with the help of $\delta l^{-2} = \delta_1 l_1^{-2}$ and $\delta l^{-5} = \delta_1 l_1^{-5}$, we find that the precession of component $m_1$ relative to the center of mass is exactly $m_2 /(m_1 + m_2)$ times that of the relative motion. Similarly, the precession of component $m_2$ relative to the center of mass is $m_1 /(m_1 + m_2)$ times that of the relative motion. Therefore, the total precession observed from the center of mass is consistent with the result of relative motion.

We now turn to the gravity effect. The metric for relative motion is [4]:

$$\mathrm{d}s^2 = c^2[1 - \frac{2G(m_1 + m_2)}{rc^2}]\mathrm{d}t^2 - [1 - \frac{2G(m_1 + m_2)}{rc^2}]^{-1}\mathrm{d}r^2 - r^2 \mathrm{d}\theta^2 - r^2 \sin^2 \theta \mathrm{d}\varphi^2 \tag{44}$$

The precession value is $6\pi G(m_1 + m_2) / ac^2(1 - e^2)$. Now we observe the motion of the binary from the center of mass. For the component star $m_1$, according to Eq. (43), we may think that there is a gravitational source with the mass of $m_2^3 /(m_1 + m_2)^2$ at the center of mass while the other component star does not exist. Similarly, for the component star $m_2$, we may think that there is a gravitational source with the mass of $m_1^3 /(m_1 + m_2)^2$ at the center of mass and the other component star does not exist. In this way, the two-body motion of the binary can be transformed into two single-body motions with respect to the center of mass. The metric of the component $m_1$ relative to the center of mass is:

$$\mathrm{d}s^2 = c^2(1 - \frac{2GM_1'}{r_1 c^2})\mathrm{d}t^2 - (1 - \frac{2GM_1'}{r_1 c^2})^{-1}\mathrm{d}r_1{}^2 - r_1{}^2 \mathrm{d}\theta^2 - r_1{}^2 \sin^2 \theta \mathrm{d}\varphi^2 \tag{45}$$

where $M_1' = m_2^3 /(m_1 + m_2)^2$. The precession is $6\pi G M_1' / a_1 c^2 (1 - e_1^2)$. Similarly, the precession of the component star $m_2$ is $6\pi G M_2' / a_2 c^2 (1 - e_2^2)$, where $M_2' = m_1^3 /(m_1 + m_2)^2$. Notice that $m_1 a_1 = m_2 a_2$, $a_1 + a_2 = a$, $e_1 = e_2 = e$, the precession of $m_1$ is $m_2^2 /(m_1 + m_2)^2$ times that of relative motion, and for component star $m_2$ the precession is $(m_1^2 + m_2^2)/(m_1 + m_2)^2$ times that of relative motion. So the total precession is $(m_1^2 + m_2^2)/(m_1 + m_2)^2$ times that of relative motion. In the case of $m_1 = m_2$, the precession observed from the center of mass is just half that of relative motion.

If inertial frame of center of mass is adopted, the difference between the observed values and the theoretical ones of the precession of binary star systems will decrease. For example, for the DI Herculis binary system ($m_1 = 5.15 M_\odot$, $m_2 = 4.52 M_\odot$, $M_\odot$ is the solar mass), the precession caused by the classical precession and gravity effect for the relative motion are 1.93°/100 years and 2.34°/100 years, respectively, and the total precession observed is $1.30 \pm 0.14$°/100 years [5]. According to our theory, the precession is 3.1°/100 years. It can be seen that even if our result has reduced the theoretical value, it still cannot match the observed value. However, according to the analysis of Ref. [6-8], if the rotation axes of the binary stars are not perpendicular to the orbital plane, the theoretical value will decrease further.

The research for 62 binary systems in Ref. [9] indicated that according to the present theory, there is an asymmetry between the situation where the theoretical value is higher than the observed one and the opposite situation: Within the error limit, there are 28 systems whose theoretical values are larger than the observed ones; while there are 14 systems for the opposite case. In the case of twice the error, there are 22 systems whose theoretical values are larger than the observed ones; while there are 7 systems for the opposite case. In the case of triple the error, there are 15 systems whose theoretical values are larger than the observed ones; while there are only 2 systems for the opposite case. For a subsample of 26 systems with the most reliably determined eccentricities and apsidal periods, this difference also exists: In the case of twice the error, there are 9 systems whose theoretical values are larger than the observed ones; while not one system is found for the opposite case.

This asymmetry can be greatly reduced by adopting the inertial frame of center of mass. The error data are taken from [9], and the masses of the components of the binary systems are taken from [10]. The results are as follows: Within the error limit, there are 26 systems whose theoretical values are larger than the observed ones; while there are 19 systems for the opposite case. In the case of twice the error, there are 17 systems whose theoretical values are larger than the observed ones; while there are 13 systems for the opposite case. In the case of triple the error, there are 15 systems whose theoretical values are larger than the observed ones; while there are 2 systems for the opposite case. For the chosen subsample of 26 systems, in the case of twice the error, there are 7 systems whose theoretical values are larger than the observed ones, and 4 systems for the opposite case. It can be seen that the asymmetry still exists after adopting the inertial frame of center of mass, which can be explained by supposing that the spin axes of the components of some binary systems are not perpendicular to the orbital planes.

## 3 Dynamical Gravitational Field

We know that a static charge generates only electrostatic field, while a moving charge also generates magnetic field. Combining Coulomb's law and special relativity, we can derive the expression of the magnetic field. The detailed method may refer to [11]. Here we briefly introduce the idea: Suppose there is a charge $Q$ moving at a constant speed $u$ relative to a still frame S, and another charge $q$ moves at a speed $v$ relative to S. We first write the interaction of $Q$ on $q$ in the frame S' that moves with the charge $Q$. The interaction is only the Coulomb force. Then according to the transformation formulas of the forces between inertial frames, we get the interaction of $Q$ on $q$ in the frame S, which includes two forces, one is electric force and the other is magnetic force.

We wish to treat the law of universal gravitation in the same way. But there are no synchronous clocks and unified rulers in the gravitational field, Lorentz transformation cannot be applied to large distances, this method fails. Nevertheless, we can apply local Lorentz transformation in the gravitational field. As mentioned earlier, by adjusting the clocks and rulers in the gravitational field, we can make them be respectively consistent with the clock and ruler in the inertial frame. Then applying the local Lorentz transformation in the gravitational field, the new metric expression will contain such terms as $\mathrm{d}t\mathrm{d}x$, $\mathrm{d}t\mathrm{d}y$, $\mathrm{d}t\mathrm{d}z$, which are the corresponding components of gravitational vector potential.

Since the Lorentz transformation is convenient to apply in the rectangular coordinate system, we need to find out the metric expression in the rectangular coordinate system. Suppose $r = r_1 f(r_1)$, then we have $\mathrm{d}r = \mathrm{d}r_1[f(r_1) + r_1 f'(r_1)]$, where $f(r_1)$ is determined by the following equation:

$$e^{\frac{2GM}{r_1 f(r_1) c^2}} = \left[\frac{f(r_1)}{f(r_1) + r_1 f'(r_1)}\right]^2 \tag{46}$$

Then we get

$$\begin{aligned}
\mathrm{d}s^2 &= c^2 e^{-\frac{2GM}{rc^2}} \mathrm{d}t^2 - e^{\frac{2GM}{rc^2}} \mathrm{d}r^2 - r^2 \mathrm{d}\theta^2 - r^2 \sin^2\theta \mathrm{d}\varphi^2 \\
&= c^2 e^{-\frac{2GM}{r_1 f(r_1) c^2}} \mathrm{d}t^2 - \left[\frac{f(r_1)}{f(r_1) + r_1 f'(r_1)}\right]^2 \left[f(r_1) + r_1 f'(r_1)\right]^2 \mathrm{d}r_1^2 - r_1^2 f^2(r_1)\left[\mathrm{d}\theta^2 + \sin^2\theta \mathrm{d}\varphi^2\right] \\
&= c^2 e^{-\frac{2GM}{r_1 f(r_1) c^2}} dt^2 - f^2(r_1)\left[\mathrm{d}r_1^2 + r_1^2 \mathrm{d}\theta^2 + r_1^2 \sin^2\theta \mathrm{d}\varphi^2\right]
\end{aligned} \tag{47}$$

Let $x = r_1 \sin\theta\cos\varphi$, $y = r_1 \sin\theta\sin\varphi$, $z = r_1 \cos\theta$, we find

$$\begin{cases}
\mathrm{d}x = \sin\theta\cos\varphi \mathrm{d}r_1 + r_1\cos\theta\cos\varphi \mathrm{d}\theta - r_1 \sin\theta\sin\varphi \mathrm{d}\varphi \\
\mathrm{d}y = \sin\theta\sin\varphi \mathrm{d}r_1 + r_1\cos\theta\sin\varphi \mathrm{d}\theta + r_1 \sin\theta\cos\varphi \mathrm{d}\varphi \\
\mathrm{d}z = \cos\theta \mathrm{d}r_1 - r_1 \sin\theta \mathrm{d}\theta
\end{cases} \tag{48}$$

Then we have $\mathrm{d}x^2 + \mathrm{d}y^2 + \mathrm{d}z^2 = \mathrm{d}r_1^2 + r_1^2 \mathrm{d}\theta^2 + r_1^2 \sin^2\theta \mathrm{d}\varphi^2$. Substituting it into (47), we obtain the metric in rectangular coordinate frame

$$\mathrm{d}s^2 = c^2 e^{-\frac{2GM}{r_1 f(r_1) c^2}} \mathrm{d}t^2 - f^2(r_1)(\mathrm{d}x^2 + \mathrm{d}y^2 + \mathrm{d}z^2) \tag{49}$$

The remaining problem is to solve for the expression of $f(r_1)$ using Eq. (46). Then we can obtain the relation between $r$ and $r_1$ with $r = r_1 f(r_1)$. The analytical solution of Eq. (46) is difficult to find. But it is easy for Schwarzschild metric. We make the following transformations [12]:

$$\begin{cases} r = r_1\left(1+\dfrac{GM}{2r_1c^2}\right)^2 \\ x = r_1\sin\theta\cos\varphi \\ y = r_1\sin\theta\sin\varphi \\ z = r_1\cos\theta \end{cases} \tag{50}$$

Expressing $r$ and $\mathrm{d}r$ in Eq. (8) with $r_1$ and $\mathrm{d}r_1$, respectively, we obtain the Schwarzschild metric in the rectangular coordinate system [12]:

$$\mathrm{d}s^2 = c^2(1-\frac{2GM}{rc^2})\mathrm{d}t^2 - (1-\frac{2GM}{rc^2})^{-1}\mathrm{d}r^2 - r^2\mathrm{d}\theta^2 - r^2\sin^2\theta\mathrm{d}\varphi^2$$

$$= \left(\frac{1-\dfrac{GM}{2r_1c^2}}{1+\dfrac{GM}{2r_1c^2}}\right)^2 c^2\mathrm{d}t^2 - \left(1+\frac{GM}{2r_1c^2}\right)^4\left(\mathrm{d}x^2+\mathrm{d}y^2+\mathrm{d}z^2\right)$$

$$\approx \left(1-\frac{2GM}{r_1c^2}\right)c^2\mathrm{d}t^2 - \left(1+\frac{2GM}{r_1c^2}\right)\left(\mathrm{d}x^2+\mathrm{d}y^2+\mathrm{d}z^2\right) \tag{51}$$

Suppose a particle $m$ moves along the $x$ axis at a velocity $v$ in the gravitational field generated by the gravitational source $M$. According to the local Lorentz transformation, we have

$$\mathrm{d}t = \frac{\mathrm{d}t' + \dfrac{v}{c^2}\mathrm{d}x'}{\sqrt{1-\beta^2}}, \quad \mathrm{d}x = \frac{\mathrm{d}x' + v\mathrm{d}t'}{\sqrt{1-\beta^2}}, \quad \mathrm{d}y = \mathrm{d}y', \quad \mathrm{d}z = \mathrm{d}z' \tag{52}$$

Substituting them into Eq. (51) yields

$$\mathrm{d}s^2 = c^2\left(1-\frac{2GM}{r_1c^2}\right)\left(\frac{\mathrm{d}t' + \dfrac{v}{c^2}\mathrm{d}x'}{\sqrt{1-\beta^2}}\right)^2 - \left(1+\frac{2GM}{r_1c^2}\right)\left[\left(\frac{\mathrm{d}x' + v\mathrm{d}t'}{\sqrt{1-\beta^2}}\right)^2 + \mathrm{d}y'^2 + \mathrm{d}z'^2\right] \tag{53}$$

In the case of $\beta << 1$, the above expression can be simplified as

$$\mathrm{d}s^2 \approx c^2\left(1-\frac{2GM}{r_1c^2}\right)\mathrm{d}t'^2 - \left(1+\frac{2GM}{r_1c^2}\right)\left[\mathrm{d}x'^2 + \mathrm{d}y'^2 + \mathrm{d}z'^2\right] - \frac{8GMv}{r_1c^2}\mathrm{d}t'\mathrm{d}x' \tag{54}$$

It should be noted that the metric expression of Eq. (51) holds only for the static gravitational field. When the gravitational source moves, its metric expression can be generally written as [12]:

$$\mathrm{d}s^2 = g_{00}c^2\mathrm{d}t^2 + 2g_{0i}\mathrm{d}t\mathrm{d}x^i + g_{ik}\mathrm{d}x^i\mathrm{d}x^k \tag{55}$$

We know that there are scalar potential and vector potential in electromagnetic theory. Similarly, there are scalar potential and vector potential in the gravitational theory. The scalar gravitational potential is $\chi = -\frac{c^2}{2}(1 - g_{00})$ and the vector gravitational potential is $A_i = -g_{0i} / \sqrt{g_{00}}$ [12]. Comparing (54) with (55), we get

$$A'_x = \frac{4GMv}{r_1 c^2 \sqrt{1 - 2GM / r_1 c^2}}, \quad A'_y = 0, \quad A'_z = 0 \tag{56}$$

which is the gravitational vector potential experienced by a test particle moving in the static gravitational field. Since the motion is relative, we may think this gravitational vector potential is generated by a gravitational source moving along the $x$ axis at the velocity of $-v$. In the case of weak gravitational field, we have $r_1 \approx r$, $GM / r_1 c^2 << 1$, so when the gravitational source moves along the $x$ axis at the velocity of $v$, the components of the gravitational vector potential are:

$$A_x = -\frac{4GMv}{rc^2}, \quad A_y = 0, \quad A_z = 0 \tag{57}$$

respectively, which agree with the results obtained by the post Newton approximation in general relativity [1]. Compared with electromagnetic vector potential, the gravitational vector potential has an additional factor of 4, which may be due to time dilation and length contraction in the gravitational field. It can be seen from Eq. (54) that the metric is symmetrical with respective to $x$, $y$ and $z$, so there will be a gravitational vector potential in the direction that the source moves, while the transverse components of the vector potential are zero.

Now we apply the above results to investigate the motion of a body in the dynamic gravitational field. We take the precession of the spin axis of the gyroscope rotating around the earth as an example. This experiment is also a well-known experiment to test general relativity. In 2004, the Gravity Probe B (GP-B) satellite was launched in America. There were four high-precision gyroscopes, which were suspended in vacuum by means of electric suspension and rotated at high speed. When the direction of the moment experienced by the gyroscope is inconsistent with the direction of the spin axis, there will be a precession of the spin axis. It is very difficult to understand and calculate the precession of the gyroscope according to general relativity, so we directly apply our theory to calculate, and in the end, we will find that our results are the same as those of general relativity.

From the classical electromagnetic theory, we know that a charged sphere will generate a magnetic moment $\boldsymbol{\mu}$ as it spins. When it is placed in an external magnetic field $\mathbf{B}$, it will experience a moment $\mathbf{M} = \boldsymbol{\mu} \times \mathbf{B}$. When the directions of $\mathbf{M}$ and $\boldsymbol{\mu}$ are parallel, the magnitude of the angular momentum will change; while when the direction of $\mathbf{M}$ is perpendicular to that of $\boldsymbol{\mu}$, the magnitude of the angular momentum remains unchanged, while the direction of the angular momentum will change, that is, it will rotate around the original direction, which is the precession of the sphere. Similarly, a high-speed rotating gyroscope will precess in the gravitomagnetic field. The GP-B experiment was aimed to measure the precession of the gyroscope caused by the gravitomagnetic field generated by the earth's revolution and spin. As general relativity does not take gravity as a force, the precessions of the gyroscope caused by the earth's revolution and spin are called geodesic effect and frame-dragging effect, respectively.

The gyroscope rotates around the earth with the satellite, so it is a non-inertial system. In order to calculate its precession relative to the inertial frame (distant star), the Thomas precession must be taken into account. Thomas precession is a purely kinematic effect. To derive Thomas precession, continuous Lorentz transformations are needed. When an object moves along a curve, two successive Lorentz transformations are equivalent to a Lorentz transformation from the start point to the end point plus a Thomas rotation. For ease of understanding, let's start with the situation of electromagnetic field. Suppose a charged particle rotates with respect to a laboratory inertial frame. The charged particle's rest frame of coordinate is defined as a co-moving sequence of inertial frames whose successive origins move at each instant with the velocity of the charged particle. The total time rate of the spin with respect to the laboratory inertial frame, or more generally, any vector $\mathbf{G}$ is given by the well-known result [13]:

$$\left(\frac{d\mathbf{G}}{dt}\right)_{\text{inertial frame}} = \left(\frac{d\mathbf{G}}{dt}\right)_{\text{rest frame}} + \boldsymbol{\omega}_T \times \mathbf{G} \tag{58}$$

where $\boldsymbol{\omega}_T$ is the angular velocity of rotation found by Thomas.

$$\boldsymbol{\omega}_T = \frac{\gamma^2}{\gamma+1}\frac{\mathbf{a}\times\mathbf{v}}{c^2} \approx \frac{1}{2}\frac{\mathbf{a}\times\mathbf{v}}{c^2} \tag{59}$$

where $\gamma = 1/\sqrt{1-\mathbf{v}^2/c^2}$, $\mathbf{a}$ is the acceleration of the particle, and $\mathbf{v}$ the rotational velocity. In the charged particle's rest frame, the equation of motion of the spin is:

$$\left(\frac{d\mathbf{J}}{dt}\right)_{\text{rest frame}} = \boldsymbol{\mu}\times\mathbf{B}' \tag{60}$$

where $\boldsymbol{\mu}$ is the magnetic moment of the charged particle, $\mathbf{B}'$ the magnetic induction intensity in the charged particle's rest frame. The classical relation between $\boldsymbol{\mu}$ and angular momentum $\mathbf{J}$ is

$$\boldsymbol{\mu} = \frac{q}{2m}\mathbf{J} \tag{61}$$

Then the motion equation of the spin of the charged particle with respect to the laboratory inertial frame is

$$\frac{d\mathbf{J}}{dt} = \boldsymbol{\mu}\times\mathbf{B}' + \boldsymbol{\omega}_T\times\mathbf{J} = \frac{q}{2m}\times\mathbf{J}\times\mathbf{B}' + \boldsymbol{\omega}_T\times\mathbf{J} = \left(-\frac{q}{2m}\times\mathbf{B}' + \boldsymbol{\omega}_T\right)\times\mathbf{J} \tag{62}$$

According to the similarity between gravity and electromagnetism, we replace charge $q$ with mass $m$, and $\mathbf{B}'$ with $\mathbf{B}'_g$, where $\mathbf{B}'_g$ is the gravitomagnetic intensity observed from the rest frame of the gyroscope. As viewed from the rest frame of the gyroscope, the earth not only moves around the gyroscope but also rotates about its spin axis. Accordingly $\mathbf{B}'_g = \nabla \times \mathbf{A}'_g$ consists of two terms: geodetic term and frame-dragging term.

We first see the geodetic term, as shown in Fig. 3. Suppose the gyroscope moves along the direction of $y$ axis, then as viewed from the gyroscope, the earth is moving in the $-y$ direction, so the components of vector potential experienced by the gyroscope are $A'_{gx} = A'_{gz} = 0$, $A'_{gy} = 4GMv/xc^2$, respectively. Then we have

$$\mathbf{B}'_g = \nabla \times A'_g = \begin{vmatrix} i & j & k \\ \partial/\partial x & \partial/\partial y & \partial/\partial z \\ 0 & 4GMv/xc^2 & 0 \end{vmatrix} = \left\{0,0,\left(\frac{4GMv}{xc^2}\right)'_x\right\} = \left\{0,0,-\frac{4GMv}{x^2c^2}\right\} \tag{63}$$

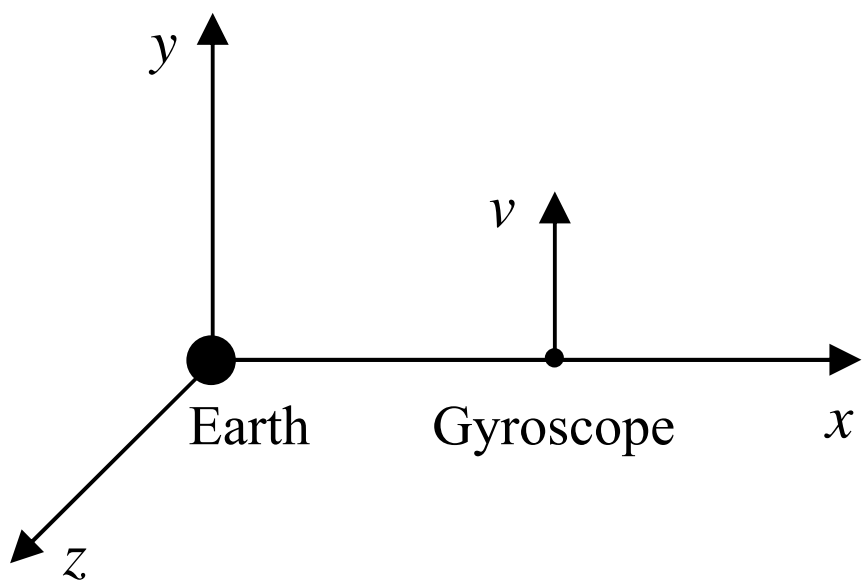

Fig. 3 The relative motion between the earth and the gyroscope.

It can be seen that the direction of $\mathbf{B}'_g$ is along the $-z$ axis. As viewed from the earth, the direction of $\mathbf{a}\times\mathbf{v}$ of the gyroscope is also in the direction of $-z$, and its magnitude is $GMv/x^2$, so the geodetic precession is:

$$\frac{d\mathbf{J}}{dt} = \left(-\frac{1}{2}\times\mathbf{B}'_g + \boldsymbol{\omega}_T\right)\times\mathbf{J} = \left(-\frac{1}{2}\frac{4\mathbf{a}\times\mathbf{v}}{c^2} + \frac{1}{2}\frac{\mathbf{a}\times\mathbf{v}}{c^2}\right)\times\mathbf{J} = -\frac{3}{2}\frac{\mathbf{a}\times\mathbf{v}}{c^2}\times\mathbf{J} \tag{64}$$

Thus the geodetic precession is $-3\mathbf{a}\times\mathbf{v}/2c^2$, which agrees with the result of general relativity [1]. It can be seen that if the gravitomagnetic potential, like the electromagnetic potential, does not have the factor of 4, then the above expression will be zero. Similarly, we can find out the influence of the spin of the earth on the precession of the gyroscope. The method is similar to that of the electromagnetic field, except that the gravitomagnetic potential has an additional factor of 4. The gravitomagnetic potential generated by the earth's spin is [1]:

$$A_g = \frac{2G}{r^3c^2}(\mathbf{x}\times\mathbf{J}_\oplus) \tag{65}$$

where $\mathbf{J}_\oplus$ is the angular moment of the earth. For the precession of the gyroscope caused by the spin of the earth, we have

$$\frac{\mathrm{d}\mathbf{J}}{\mathrm{d}t} = \frac{m}{2m}\mathbf{J}\times\mathbf{B}'_g = -\frac{1}{2}\mathbf{B}'_g\times\mathbf{J} = -\frac{1}{2}\nabla\times A_g\times\mathbf{J} \tag{66}$$

So the precession angular velocity of the gyroscope is $-\frac{1}{2}\nabla\times A_g$. As the angle between $\mathbf{x}$ and $\mathbf{J}_\oplus$ changes with the position of the gyroscope, the average precession angular velocity is $\frac{GI\omega_e}{2r^3c^2}$ [1], where $I$ is the moment of inertia and $\omega_e$ the spin angular velocity of the earth.

The GP-B satellite was launched on April 20, 2004. Data collection began in August and lasted for one year, while the data analysis work lasted for six years, and the final experimental results were published in May 2011 [14]. GP-B satellite flies in a circular polar orbit 642km above the ground. The spin axis of the gyroscope is parallel to the equatorial plane. Thus the geodesic precession is in the North-South direction, and the frame-dragging precession is in the West-East direction. The two precessions are perpendicular to each other and can be measured separately. In terms of (60), the theoretical value of geodesic precession is 6606.1 mas/yr, and the experimental value is $6601.8 \pm 18.3$ mas/yr. The frame-dragging effect is much smaller. With $I = 7.7\times10^{37}\,\mathrm{kgm}^2$ and $\omega_e = 7.29\times10^{-5}$ rad/s, we find that the theoretical value of frame-dragging precession is 39.2 mas/yr, while the experimental value is $37.2 \pm 7.2$ mas/yr.

Note that since the sun is an accurate inertial frame in the solar system, the geodesic effect caused by the gyroscope rotating around the sun with the earth should also be taken into account. Simple calculation shows that the precession is about 19 mas/yr. Because the angle between the equatorial plane and the orbital plane of the earth is $\theta = 23.5^\circ$, only a fraction of $19\cos\theta$ can lead to the precession of the gyroscope, and it can be decomposed into the precession in the West-East direction ($19\cos\theta\cos\theta$) and North-South direction ($19\cos\theta\sin\theta$). Therefore, the main contribution is in the West-East direction. One may see [15] for detailed discussion. In contrast, the influence of the moon on the gyroscope is much smaller and can be ignored.

## 4 Gravitational Radiation

### 4.1 The theory of gravitational radiation

We know that an accelerated electrical charge will emit electromagnetic waves. Similarly, an accelerated body will emit gravitational waves. For an isolated charge system, the strongest radiation is electric dipole moment radiation. The dipole moment is

$$\mathbf{d}_e = \sum_i e_i \mathbf{r}_i \tag{67}$$

where $e_i$ and $\mathbf{r}_i$ are the charge and the position vector of particle $i$, respectively. The radiant intensity of dipole moment is proportional to $\ddot{\mathbf{d}}_e$. If we replace $e_i$ with $m_i$, we will obtain the mass dipole moment of an isolated system

$$\mathbf{d}_m = \sum_i m_i \mathbf{r}_i \tag{68}$$

whose first order derivative is the total momentum of the system:

$$\dot{\mathbf{d}}_m = \sum_i m_i \dot{\mathbf{r}}_i \tag{69}$$

Since the total momentum of an isolated system is conserved, we have $\ddot{\mathbf{d}}_m = \dot{\mathbf{p}} = 0$. Thus the mass dipole moment radiation does not exist in gravity physics. The second strongest radiations are magnetic dipole moment radiation and electric quadrupole moment radiation, respectively. The radiant intensity of magnetic dipole moment is determined by its second order derivative. The magnetic dipole moment is:

$$\boldsymbol{\mu} = \frac{1}{2}\sum_i \mathbf{r} \times (e_i \mathbf{v}_i) \tag{70}$$

Replacing $e_i$ with $m_i$, we obtain the gravitomagnetic dipole moment

$$\boldsymbol{\mu}_g = \frac{1}{2}\sum_i \mathbf{r} \times (m_i \mathbf{v}_i) \tag{71}$$

which is just half of the angular momentum of the system. Due to the conservation of the angular momentum of an isolated system, there does not exist gravitomagnetic dipole moment radiation. The gravitational radiation similar to electric quadrupole moment radiation does exist. For an isolated system, the main gravitational radiation is mass quadrupole moment radiation. The mass quadrupole moment is

$$\mathbf{D}_{\alpha\beta} = \int \rho(3x^\alpha x^\beta - \delta_\alpha^\beta x^\gamma x^\gamma) d^3 x \tag{72}$$

The total power radiated is

$$\frac{dE}{dt} = -\frac{G}{45c^5}\dddot{\mathbf{D}}_{\alpha\beta}^2 \tag{73}$$

The above simple discussion can refer to [16]. More detailed discussions may see [1, 12]. Our theory is equivalent to general relativity in the case of weak field and low speed. Thus the two give a same result for the gravitational radiation of isolated system.

**4.2 The gravitational radiation of binary**

For the gravitational radiation of binary star system, if viewed from the center of mass of the system, the result will be different from the present one based on the relative motion of binary. First, let's consider the circular motion of a point particle, whose gravitational radiation is [1]

$$-\frac{\mathrm{d}E}{\mathrm{d}t} = \frac{32}{5}\frac{G}{c^5}\omega^6 m^2 r^4 \tag{74}$$

Now generalize the above result to the relative motion of the binary, then $m$ should be substituted for the reduced mass $m_1 m_2 /(m_1 + m_2)$. The angular frequency of the binary is $\omega^2 = G\frac{(m_1 + m_2)}{R^3}$, where $R$ is the distance between the two components. Then the quadrupole radiation of the binary is [12]

$$-\frac{\mathrm{d}E}{\mathrm{d}t} = \frac{32}{5}\frac{G^4}{c^5} m_1^2 m_2^2 (m_1 + m_2) / R^5 \tag{75}$$

Now we observe the motion of the binary from the center of mass of the system. In this case, the total radiation power is the sum of the radiation power of the two components, then we have

$$-\frac{\mathrm{d}E}{\mathrm{d}t}=\frac{32}{5}\frac{G}{c^5}\omega^6 m_1^2 R_1^4+\frac{32}{5}\frac{G}{c^5}\omega^6 m_2^2 R_2^4=\frac{32}{5}\frac{G^3(m_1^2R_1^4+m_2^2R_2^4)(m_1+m_2)^3}{(R_1+R_2)^9c^5} \tag{76}$$

In the case of $m_1=m_2$ and $R_1=R_2=R/2$, the above result is only half that of Eq. (75). Now consider the motion of the binary in elliptical orbits. In general relativity, the radiation power is the result of (75) multiplied by a factor of $f(e)$, and $R$ should be replaced by the semimajor axis of the relative motion orbit of the binary [12], that is

$$-\frac{\mathrm{d}E}{\mathrm{d}t}=\frac{32}{5}\frac{G^4}{c^5}\frac{m_1^2m_2^2(m_1+m_2)}{a^5}f(e) \tag{77}$$

The expression of $f(e)$ is

$$f(e)=\frac{1+\frac{73}{24}e^2+\frac{37}{96}e^4}{(1-e^2)^{7/2}} \tag{78}$$

where $e$ is eccentricity. Since the center of mass is an inertial frame, the radiation power of the binary should be

$$-\frac{dE}{dt}=\frac{32}{5}\frac{G^4}{c^5}\frac{(m_1^2a_1^4+m_2^2a_2^4)(m_1+m_2)^3}{(a_1+a_2)^9}f(e) \tag{79}$$

where $a_1$ and $a_2$ the semimajor axes of the components $m_1$ and $m_2$ with respect to the center of mass, respectively. The above result is less than that of (77).

## 5 The Mass of Neutron Star

In order to study the mass of neutron star, it is necessary to establish the hydrostatic equilibrium equation in the gravitational field. First, we consider the simplest situation, i.e. the static equilibrium of fluid in the Newtonian mechanics. As shown in Fig. 4, there is a volume element $\mathrm{d}V=\mathrm{d}s\mathrm{d}r$ with the mass of $\mathrm{d}m=\rho\mathrm{d}V$ at the radius $r$ in the star. The pressure difference between the lower surface and the upper surface of the volume element is $\mathrm{d}F=[p(r)-p(r+\mathrm{d}r)]\mathrm{d}s=-\mathrm{d}p\mathrm{d}s$, where $p(r)$ is the pressure at radius $r$. $\mathrm{d}F$ balances with gravity at $r$, which follows

$$\mathrm{d}F=-\mathrm{d}p\mathrm{d}s=\frac{GM(r)}{r^2}\mathrm{d}m=\frac{GM(r)}{r^2}\rho(r)\mathrm{d}s\mathrm{d}r \tag{80}$$

where $M(r)$ is the mass contained in the sphere with a radius of $r$. Simplifying the above equation yields

$$\frac{\mathrm{d}p}{\mathrm{d}r}=-\frac{GM(r)}{r^2}\rho(r) \tag{81}$$

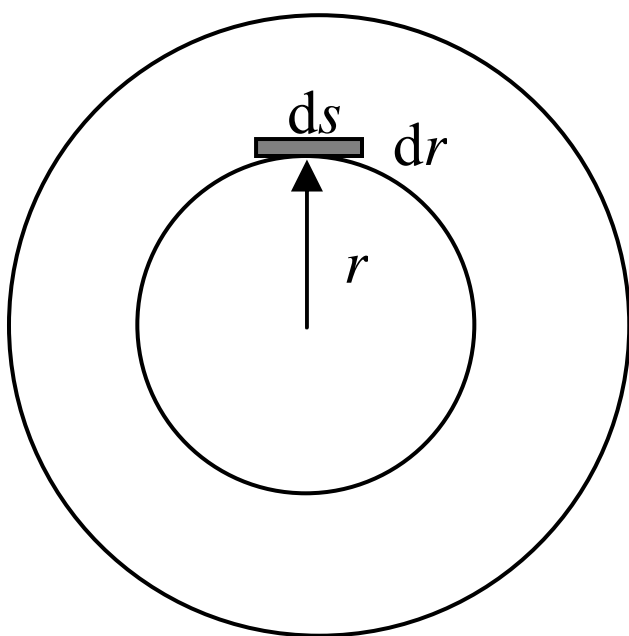

Fig. 4 The hydrostatic equilibrium in a star.

The relativistic hydrostatic equilibrium equation is similar to the above equation, with the exception that $\rho(r)$ is substituted for $\rho(r)+p(r)/c^2$, i.e.,

$$\frac{\mathrm{d}p}{\mathrm{d}r}=-\frac{GM(r)}{r^2}(\rho+p/c^2) \tag{82}$$

Now consider the gravity effect on the above equation. According to the previous discussion, we know that the law of universal gravitation should be modified as $F=GMm\mathrm{e}^{GM/rc^2}/r^2$. We have two different understandings of this formula: Taking the body $M$ as the reference frame, we may think that the mass of a particle $m$ will increase when it falls freely in the gravitational field, so the factor $\mathrm{e}^{GM/rc^2}$ is caused by the conversion of potential energy into kinetic energy. When observed from the co-moving coordinate frame of particle $m$, this factor is caused by the energy variation of the virtual gravitons in the gravitational field. In general relativity, $\rho$ and $p$ are defined in the co-moving frame, that is, they are proper density and proper pressure, respectively. Thus when considering the gravity effect, we should make the following substitutions: $M(r)\to M(r)\mathrm{e}^{GM/rc^2}$, $r\to r\mathrm{e}^{GM/rc^2}$, $\mathrm{d}r\to \mathrm{d}r\mathrm{e}^{GM/rc^2}$. Then we will find that the hydrostatic balance equation in the gravitational field is the same as that without the gravity. As we have mentioned earlier, the gravitational field exerts the same influence on the physical process and the space-time. The hydrostatic equilibrium equation is just an example; it is the same with and without gravity.

Now let's consider the effect of gravity on the hydrostatic equilibrium equation in another way. The general form of static spherically symmetric metric is [1]:

$$\mathrm{d}s^2=-a(r)\mathrm{d}r^2-r^2\mathrm{d}\theta^2-r^2\sin^2\theta\mathrm{d}\varphi^2+b(r)\mathrm{d}t^2 \tag{83}$$

For Schwarzschild exterior solution, we have $a(r)=(1-2GM/rc^2)^{-1}$ and $b(r)=(1-2GM/rc^2)$. For Schwarzschild interior solution, we still have $a(r)=(1-2GM(r)/rc^2)^{-1}$, but $b(r)\neq(1-2GM/rc^2)$ [1]. According to our interpretation of gravity, the gravitational field at radius $r$ inside a spherically symmetric star is only related to $M(r)$, and $M(r)$ can be regarded as concentrated at the center of the star, so the forms of metric expressions inside and outside the star are the same, so we have $a(r)=\mathrm{e}^{2GM(r)/rc^2}$, $b(r)=\mathrm{e}^{-2GM(r)/rc^2}$. The hydrostatics equilibrium equation in general relativity is [1]:

$$\frac{b'}{b} = -\frac{2p'}{\rho + p/c^2} \tag{84}$$

where $b' = \mathrm{d}b/\mathrm{d}r$. Substituting $b(r) = \mathrm{e}^{-2GM(r)/rc^2}$ into above equation just leads to (82). Therefore, even in general relativity, the hydrostatic equilibrium equation is still (82) as long as $b(r) = \mathrm{e}^{-2GM(r)/rc^2}$ holds. As $b(r) \neq (1 - 2GM/rc^2)$ in general relativity, the hydrostatic equilibrium equation inside a star has a strange form:

$$\frac{\mathrm{d}p}{\mathrm{d}r} = -\frac{GM(r)}{r^2}(\rho + \frac{p}{c^2})(1 + \frac{4\pi r^3 p}{M(r)c^2})(1 - \frac{2GM(r)}{rc^2})^{-1} \tag{85}$$

which is Tolman-Oppenheimer-Volkoff equation (TOV equation). Compared with (82), it has two more factors. The reason that it is different from (82) is that Einstein's gravitational field equation is of second order. From the exact metric expression of (7), it can be seen that the gravitational field equation is of infinite order. The assumption of second-order gravitational field equation not only brings about the problem of black holes and singularities, but also leads to the inconsistency of the forms of Schwarzschild exterior and interior solutions.

The differential equation for $M(r)$ is:

$$\frac{\mathrm{d}M(r)}{\mathrm{d}r} = 4\pi r^2 \rho(r) \tag{86}$$

This equation is the same whether the gravity effect is considered or not. In addition, we also need to know the equation of state inside the neutron star, that is, the relation between $\rho$ and $p$. For the ideal neutron gas, the Chandrasekhar equation of state may be used [12]. Introducing a parameter,

$$t = 4\,\mathrm{In}[x + \sqrt{(1 + x^2)}] \tag{87}$$

The equation of state can be simplified as:

$$\rho = \kappa(\mathrm{sh}\,t - t)/c^2 \tag{88}$$

$$p = \frac{1}{3}\kappa(\mathrm{sh}\,t - 8\,\mathrm{sh}\frac{1}{2}t + 3t) \tag{89}$$

$$\kappa = \frac{\pi m^4 c^5}{4h^3} \tag{90}$$

where $m$ is the mass of the neutron, $h$ Planck constant, $c$ the speed of light. In terms of (82), we have

$$\frac{\mathrm{d}t}{\mathrm{d}r} = \frac{\mathrm{d}p}{\mathrm{d}r}\frac{\mathrm{d}t}{\mathrm{d}p} = -\frac{GM(r)(\rho + p/c^2)}{r^2}\frac{3}{\kappa(\mathrm{ch}t - 4\mathrm{c\,h}\frac{1}{2}t + 3)} \tag{91}$$

With (86) and (91), we can compute the mass $M$ and radius of the neutron star. The boundary conditions are:

$$r = 0\,,\quad t = t_0\,,\quad M(0) = 0$$

$$r = R\,,\quad t = 0\,,\quad M(R) = M$$

Eqs. (86) and (91) are first-order differential equations and cannot be solved analytically, but we may solve them numerically. By using of fourth-order Runge-Kutta method, if the initial values are known, the final values of the desired physical quantities can be computed with high accuracy. Different initial values of $t_0$ will give different neutron star mass and radius. The source program that is based on C language is given in appendix A, the interested readers may verify the results by themselves. The results are given in table 1.

Table 1 The computed results for the masses of neutron star

| $t_0$ | $M/M_\odot$ [a] | $M/M_\odot$ [b] | $M/M_\odot$ [c] |
|---|---|---|---|
| 1 | 0.30 | 0.30 | **0.33** |
| 2 | 0.60 | 0.61 | **0.80** |
| 3 | 0.71 | 0.70 | **1.20** |
| 4 | 0.64 | 0.66 | **1.40** |
| 5 | - | 0.55 | **1.41** |
| 8 | - | 0.37 | **0.97** |
| 10 | - | 0.42 | **0.76** |
| 15 | - | 0.42 | **0.92** |
| 18 | - | 0.43 | **0.87** |

The data in column a were computed by Oppenheimer and Volkoff; while the data in column b are computed by the author according to TOV equation; and the data in column c are computed in terms of Eqs. (86) and (91).

It should be noted that the results in table 1 hold for pure neutron stars under the ideal equation of state. The internal structure and equation of state of an actual neutron star are much more complex, especially for the core of neutron star where hyperonic fluids, $\pi$ condensation, solidification may occur. The masses of neutron stars predicted by different models range from 1.3 to 2.7 $M_\odot$, the detailed discussion may see [17]. It is particularly worth mentioning that Rhoade and Ruffini derived an upper mass limit of 3.2 $M_\odot$ starting with general elativity, causality and Le Chatelier's principle, independent of the equations of state [18]. On the basis of Rhoade and Ruffini's work, Kalogera and Baym used the modern equation of state for neutron star to fit the experimental data of nucleon-nucleon scattering and the properties of light nuclei. They found an upper mass limit of 2.9 $M_\odot$ [19]. Godzieba et al. adopted a Markov Chain Monte Carlo approach to generate about two million phenomenological equations of state with and without first-order phase transitions. They fixed the crust equation of state and only assumed causality at higher densities. The maximum mass of the neutron star they obtained was also 2.9 $M_\odot$ [20].

Now let's see the observation results of neutron star mass. Thorsett and Chakrabarty analyzed about 50 radio pulsars and found that the mass distribution of 21 radio pulsars and 5 neutron star companions are consistent with a remarkably narrow underlying Gaussian mass distribution with a peak value of $1.35 \pm 0.04$ $M_\odot$ [21]. Jim Lattimer's statistical analysis of 66 binary star systems showed that (one may refer to https://stellarcollapse.org/nsmasses.html) for 18 x-ray/optical binaries, the average mass of neutron stars is 1.71 M⊙; for 16 double neutron star binaries, the average value is 1.33 M⊙; for 29 white dwarf-neutron star binaries, the average value is 1.58 M⊙; the masses of three main sequence-neutron star binaries are between 1.58 and 1.71 M⊙. Of the 66 neutron stars, only 7 have a mass of more than 2 M⊙ and the largest mass is 2.74 M⊙. It can be seen that our theoretical results are closer to the observed ones than those of TOV equation.

Finally, it should be pointed out that the precession of periastron and the gravitational radiation of binary systems are usually used as constraints to determine the masses of binaries. However, the present equations of motion are based on relative motion. If observed from center of mass of the system, the masses of neutron stars will vary accordingly, so we need to recalculate their masses.

## 6 Black Hole and Singularity

Black hole is a very strange celestial body predicted by general relativity. Its gravity is so strong that anybody (including light) cannot escape out of it. In 1916, the German astronomer Schwarzschild derived the spherically symmetric vacuum solution of Einstein's gravitational field equation, i.e. Schwarzschild metric, as shown in (8). The coefficients in front of $\mathrm{d}t^2$ and $\mathrm{d}r^2$ must be greater than 0, which requires that $(1-2GM/rc^2)>0$. It follows that $r>2GM/c^2$, this radius is called Schwarzschild radius. Once the radius of the celestial body is less than Schwarzschild radius, the gravity inside the celestial body will become infinite. No force can resist this infinite gravity. Therefore, once a body falls into a black hole, it will finally fall to the center, forming a singularity with infinitesimal volume and infinite density. In mathematics and physics, infinity is an unacceptable thing, and the existence of singularity also violates the uncertainty principle in quantum mechanics, because the position of a point with infinitesimal volume is completely definite, so black hole and singularity should not appear in a rational physical theory. Then what's wrong with general relativity? Einstein's gravitational field equation is

$$G_{\mu\nu}=-8\pi T_{\mu\nu} \tag{92}$$

The equation itself has no problem. The problem lies in the following assumption: $G_{\mu\nu}$ is assumed to be a linear combination of the metric and its first and second order derivatives. If we expand $\mathrm{e}^{GM/rc^2}$ into series, we will find that it contains derivatives of infinite order. From the comparison between Eqs. (7) and (8), it can be seen that the real gravitational field equation contains derivatives of infinite order, so it is not surprising that the gravitational radius appears in Schwarzschild metric. This assumption not only leads to black hole and singularity, but also the inconsistency of the form of Schwarzschild exterior and interior solutions (which has been discussed in the previous section).

If black holes do not exist, what are the black holes at the centers of galaxies observed in recent years? As a matter of fact, they are little primordial universes. When the core mass of the star is greater than 3 M⊙, the degenerate pressure of neutrons can not resist the gravity inside the star, so neutrons will further break into quarks and electrons. The star will become a soup of particles containing quarks, electrons, neutrinos and other elementary particles and their antiparticles. The internal gravity of the star is balanced by the degenerate pressure of elementary particles, which is the state of the primordial universe. Although the internal temperature of the primordial universe is very high, it does not emit heat and light, so it is indistinguishable from a black hole in appearance.

## 7 Inertial Frame and Non-inertial Frame

### 7.1 The center of mass of a system as the inertial frame

Inertial frame is a frame of reference in which a body not subjected to any external force will remain at rest or move in uniform motion in a straight line. In other words, inertial frame is the frame in which Newton's first and second laws hold. But what situation can be said to be free from external force? The answer is: When a body remains at rest or moves in a uniform straight line in an inertial frame, it is not subject to external forces. Then we fall into a logical argument cycle.

As we know, special relativity applies to inertial frames of reference. However, inertial frame cannot be strictly defined in special relativity. Therefore, Einstein put forward the principle of general relativity: All reference frames are equally valid to describe the nature. But if we accept this principle, we will be immediately faced with a very serious conceptual problem: Does the earth go around the sun, or the sun goes around the earth? This is actually the argument between geocentric theory and heliocentric theory. According to modern knowledge, we know that the earth revolves around the sun. But according to the principle of general relativity, both perspectives are right. It seems that we have to introduce an impartial third party to make a judgment. The third party must jump out of the solar system and stand far away to observe the motion of the solar system. Then we will find that the heliocentric theory is correct. In fact, we don't need to do this at all. We can infer the motion of planets by observing the motion of stars in other galaxies. In galaxies, stars revolve around the center of the galaxy. Similarly, in the solar system, planets revolve around the center of the solar system (the sun).

The gravity at infinity is zero. According to the definition of inertial frame, infinity is an inertial frame. In the GP-B experiment, the binary star HR8703 (IM Pegasi) in Pegasus is employed as the guide star [14]. The reason for choosing it is that it is bright and convenient to be used as a reference frame. However, since the binary is only about 316 light-years away from the earth, the distance is not infinite, so we need to know its proper motion in the sky background, that is, the motion relative to the more distant quasar 3C454.3, which may be used as the inertial reference frame in the GP-B experiment.

The inertial frame at infinity is not convenient to use. For a system consisting of many bodies (e.g. the solar system), it moves freely as a whole in the gravitational field generated by other celestial bodies in the galaxy. According to the principle of equivalence, it is a local inertial system. For the motion of a body inside the system, what's more meaningful is its motion relative to other bodies in the system, so we need to find an inertial frame in the system. It can be proved that the center of mass of the system is an inertial reference frame. The proof has been given in many textbooks (e.g. [22]). Suppose there is a particle group consisting of $n$ particles, and the mass of the particle $i$ is $m_i$. We choose a reference frame in which the position vector of particle $i$ is $\mathbf{r}_i$. If the resultant force acting on particle $i$ is $\mathbf{F}_i$, we have

$$\frac{\mathrm{d}^2(m_i\mathbf{r}_i)}{\mathrm{d}t}=\mathbf{F}_i \qquad i=1,2,\cdots\cdots,n \tag{93}$$

There are totally $n$ particles, correspondingly there are $n$ equations. Adding these equations together yields

$$\sum_{i=1}^{n}\frac{\mathrm{d}^2(m_i\mathbf{r}_i)}{\mathrm{d}t}=\sum_{i=1}^{n}\mathbf{F}_i \tag{94}$$

If we don't consider external forces, according to Newton's third law of motion, the resultant force of all the internal forces is 0, that is,

$$\sum_{i=1}^{n}\frac{\mathrm{d}^2(m_i\mathbf{r}_i)}{\mathrm{d}t}=0 \tag{95}$$

Now we shall use the definition of the center of mass, whose expression is

$$\mathbf{r}_c=\frac{\sum_{i=1}^{n}m_i\mathbf{r}_i}{m} \tag{96}$$

where $m=\sum_{i=1}^{n}m_i$ is the total mass of the particles. Substituting (96) into (95) yields

$$\sum_{i=1}^{n}\frac{\mathrm{d}^2(m_i\mathbf{r}_i)}{\mathrm{d}t}=\frac{\mathrm{d}^2(m\mathbf{r}_c)}{\mathrm{d}t}=\frac{\mathrm{d}(m\mathbf{v}_c)}{\mathrm{d}t}=0 \tag{97}$$

Thus the velocity of the center of mass $\mathbf{v}_c$ is a constant, indicating that the center of mass is an inertial frame of reference.

Once the center of mass is chosen as the inertial frame, the previous controversy can be settled. Since there is no rotation between inertial frames, when observed from infinity, one will find that the sun does not move while the earth rotates around the sun. Strictly speaking, there is only one inertial frame of reference in the universe, that is, the center of the universe, where the big bang took place. Each galaxy revolves around the center of the universe, but the acceleration of rotation is very small, so the center of the galaxy may be used as an inertial frame when studying the motion of stars in that galaxy. Further, when studying the motion of planets in the solar system, we may take the sun as an inertial frame. And further, when we study the objects moving around the earth, such as missiles, aircrafts, satellites, we may take the geocenter as an inertial frame. Even further, when we study objects moving in a small range on the earth's surface, such as cars, ships, people, animals, we may take the ground as an inertial frame, by doing so we actually ignores the spin of the earth. When we study the motion of elementary particles in the laboratory, we often regard the laboratory as an inertial frame, because the laboratory is firmly connected with the ground.

The hypothesis of center of mass as an inertial frame can also explain Newton's rotating bucket experiment. A bucket of water is suspended by a rope. At first, both the bucket and the water remain at rest, and the water surface is flat. Then let the bucket rotate at angular velocity $\omega$. At the beginning, the water is not driven by the bucket, so the water surface is still flat. Then the friction between the bucket and the water will gradually drive the water to rotate, and the water surface becomes concave. And then let the bucket suddenly stop rotating, but the water will still rotate at the angular velocity $\omega$. At this moment, the water surface is still concave. The experimental results can be summarized as follows: When there is a relative motion between the bucket and water, the water surface can be flat or concave; when there is no relative motion between the bucket and water, the water surface can also be flat or concave. This shows that the shape of the water surface has nothing to do with the relative motion between the water and the bucket. Assuming that there is only this bucket of water in the universe, then which frame of reference is the rotation of water relative to? Newton believed that there is an absolute space in the universe, and the rotation of the water is with respect to the absolute space. According to special relativity, there is no absolute space, all motions are relative, and the inertial force is caused by the accelerated motion of the body relative to the inertial frame. But where to find the inertial frame when there are only the bucket and the water? It can be seen that neither Newtonian mechanics nor special relativity can solve the fundamental issue of inertial frame of reference. If the center of mass of the system is used as an inertial frame, the above difficulty in the rotating bucket experiment does not exist. The rotation of water and bucket is relative to their common center of mass, or relative to the vertical axis passing through the center of mass. During the rotation of the bucket and water, the rotation speed of this vertical axis is always zero, so it is an inertial frame. Then there is no need to seek for the inertial frame in outside space.

### 7.2 The frame of reference in two-body problem

For the two-body problem, we may decompose the two-body motion into two parts: the motion of the center of mass and the relative motion between the two bodies. The motion of the center of mass is free, since there is no external force. When the problem concerned depends only on the interaction between the two bodies, we may generalize the result of one-body motion to that of the two-body motion by making the following substitutions:

$$M \to m+M\ ,\ \ m \to \frac{mM}{m+M}\ ,\ \ a \to a_1 + a_2 \tag{98}$$

where $M$ is the mass of the gravitational source (e.g. the solar mass in the solar system), $m$ is the mass of the test particle, $a$ is the semi-major axis of the elliptical orbit of relative motion, while $a_1$ and $a_2$ are respectively the semi-major axes of the elliptical orbit of the two components relative to the center of mass. For example, the motion of the planet obeys Kepler's third law of motion, $a^3 / T^2 = GM / 4\pi^2$. When extended to the situation of two-body motion, it becomes $(a_1+a_2)^3/T^2=G(M+m)/4\pi^2$ according to Eq. (98). This result can also be derived by observing the motions of the two components from the center of mass of the system. For component $m$, we have

$$\frac{mv_1^2}{r_1}=\frac{GMm}{(r_1+r_2)^2} \tag{99}$$

where $r_1$ is the distance between $m$ and the center of mass, $r_2$ is the distance between $M$ and the center of mass, and we have $mr_1=Mr_2$. Substituting $r_2=mr_1/M$ into the above equation, we get

$$\frac{mv_1^2}{r_1}=\frac{GMm}{(r_1+mr_1/M)^2}=\frac{GM^3m}{r_1^2(m+M)^2} \tag{100}$$

It can be seen that the motion of component $m$ relative to the center of mass is equivalent to that there is a gravitational source with mass $M'=M^3/(m+M)^2$ at the center of mass while the other component $M$ does not exist. In this case, the motion of body $m$ is equivalent to the one-body motion, so we have $a_1=\sqrt[3]{GM'T^2/4\pi^2}$. Similarly, we have $a_2=\sqrt[3]{Gm'T^2/4\pi^2}$, where $m'=m^3/(m+M)^2$. The two expressions give the desired result $(a_1+a_2)^3/T^2=G(M+m)/4\pi^2$.

We see another example. According to Bohr's atomic theory, we have

$$\begin{cases} mvr=\hbar \\ \dfrac{mv^2}{r}=\dfrac{ke^2}{r^2} \end{cases} \tag{101}$$

where $k=1/4\pi\varepsilon_0$. From the above formulas, we obtain the energy level of the electron

$$E=\frac{1}{2}mv^2-\frac{ke^2}{r}=\frac{1}{2}\frac{k^2e^4m}{n^2\hbar^2} \tag{102}$$

In the hydrogen atom, both electron and nucleus rotate around their common center of mass, which is actually a problem of two-body motion. According to (98), the energy level of the atom is

$$E=-\frac{1}{2}\frac{k^2e^4}{n^2\hbar^2}\frac{mM}{m+M} \tag{103}$$

where $M$ is the mass of nucleus, and $Mm/(M+m)$ is the reduced mass. Now we consider the motions of the electron and the nucleus from the center of mass. In this case, the quantization condition of the electron's angular momentum should be modified to that of the atom, so we have

$$\begin{cases} m_1v_1r_1+m_2v_2r_2=n\hbar \\ m_1r_1=m_2r_2 \\ \dfrac{m_1v_1^2}{r_1}=\dfrac{ke^2}{(r_1+r_2)^2} \\ \dfrac{m_2v_2^2}{r_2}=\dfrac{ke^2}{(r_1+r_2)^2} \end{cases} \tag{104}$$

The above equations include four unknown quantities $v_1$, $v_2$, $r_1$, $r_2$. After simple calculations, we obtain

$$\begin{cases} v_1 = \frac{ke^2}{n\hbar}\frac{m_2}{m_1+m_2} \\ v_2 = \frac{ke^2}{n\hbar}\frac{m_1}{m_1+m_2} \\ r_1 = \frac{n^2\hbar^2}{ke^2}\frac{1}{m_1} \\ r_2 = \frac{n^2\hbar^2}{ke^2}\frac{1}{m_2} \end{cases} \tag{105}$$

Then the energy level of hydrogen atom is

$$E = \frac{1}{2}m_1 v_1^2 + \frac{1}{2}m_2 v_2^2 - \frac{ke^2}{r} = -\frac{1}{2}\frac{k^2e^4}{n^2\hbar^2}\frac{m_1 m_2}{m_1+m_2} \tag{106}$$

which agrees with (103). In addition, for the classical precession of binary, because the precession is only related to the interaction between the two components, the result of relative motion is the same as that of two one-body motions with respect to center of mass. But for the precession caused by gravity effect, the result is not only related to the interaction between the two components, but also related to the gravity effect, so the result of one-body motion cannot be simply extended to that of two-body motion. Similarly, the gravitational radiation of binary is related to both the interaction between the two components and gravity effect. Therefore, the generalized formula of one-body gravitational radiation does not hold for that of binary star system.

### 7.3 The superiority of inertial frame of reference

It is believed in general relativity that all reference frames are equally valid to account for the physical phenomena. However, from the above discussions, we see that in some cases, the observed results are only correct in the inertial frame. In addition, from the inertial frame, it is easier to find the law of motion. For example, when observing from the sun, we find that the orbit of the planet is an ellipse; if we observe from the earth, the orbit of the planet will be much complicated. To correctly describe the motion of a body in non-inertial frame, inertial forces must be added to the equations of motion. For example, an observer standing on the ground will find that when a body falls freely from a certain height above the ground, the landing point will be slightly east of the starting point. If viewed from the geocentric inertial frame, this is the consequence of the earth rotating eastward. Since the rotational velocity of the body is $v = r\omega$, its velocity at the starting point is larger than that of the ground. During the falling process, the initial rotational speed of the body remains unchanged in the tangential direction, so when it lands, it will run to the front of the starting point, that is, to the east. In order to explain this phenomenon in the ground reference frame, we need to introduce the Coriolis force. For the observer on the ground, this force is very hard to understand and imagine. In addition, we see that there exists a difference of Thomas precession between the inertial frame and a rotating non-inertial frame for the time rate of a vector.

To summarize, the inertial frame is indeed a superior reference frame. Therefore, to correctly describe the expansion of the universe, the center of mass of the universe must be chosen as the reference frame, as discussed below.

## 8 The Expansion of the Universe and Dark Energy

The application of the general covariance principle in cosmology is the cosmological principle, which states that at the cosmological scale, the three-dimensional space is uniform and isotropic at any time. The principle is based on the following two points:

(i) According to observations on earth, galaxy distribution, radio source count and microwave background radiation are basically uniform and isotropic within a range of no less than $10^8$ light years.

(ii) The generalization of the Copernicus principle. The heliocentric theory overturned the geocentric theory. One of its natural extensions is that the universe has no center, so the same scene of the evolution of the universe will be observed from any celestial body. Thus we can establish a universal cosmic time in the universe to investigate the evolution of the universe.

The big bang theory holds that the universe evolved from a singularity. The first question to be discussed is: What is the expansion of the universe relative to? It is believed in general relativity that the expansion of the universe is relative to the space itself. A frequently mentioned example used to describe the expansion of the universe is the relative motion of the points on the surface of an expanding balloon, where each point is far away from other points. According to our viewpoint of inertial frame of center of mass, the expansion of the universe should be relative to the center of the universe. Strictly speaking, only when observed from the center of mass is the universe isotropic. However, due to the huge scale of the universe, as long as the location is not far away from the center, the surrounding space is almost uniform and isotropic. Although our Milky way is not necessarily the center of the universe, observations show that the space around the earth is almost uniform and isotropic, so the Milky Way must be close to the center of the universe, not at the edge of the universe. For simplicity, let's assume that our space is the center of the universe.

If the universe has a boundary, maybe everyone wants to know: What's outside the universe? The answer is there is an absolute vacuum outside the universe. The degree of vacuum can be represented by the number of molecules in the space, the motion and collision of the molecules will form pressure, so the degree of vacuum can be denoted by pressure. When there is no molecule in space, the pressure is zero, that is, a complete vacuum. However, even in a complete vacuum, there will still be photons, neutrinos, gravitons and other real particles. Further, assume that we can shield these real particles, there may still be virtual particles such as virtual photons and virtual gravitons in space, which form electromagnetic and gravitational fields, respectively. As long as in the universe, there will inevitably be real particles and virtual particles. In the outermost part of the universe, the electromagnetic field and gravitational field are zero, and there are neither real particles nor virtual particles in space, which is an absolute vacuum.

If the universe has a boundary, where is the boundary of the universe? The answer is the farthest place that light can reach is the boundary of the universe. Imagine one is standing at the center of the city at night, and the lights he sees are almost the same everywhere; if he stands on the edge of the city, he would see the lights shining when he looked at the center of the city, while he would find it dark when he looks back. The similar scene might be found at the edge of the universe. Outside the universe, space has lost its meaning, so it is meaningless to discuss how large the space outside the universe is. Similarly, in a space where everything remains unchanged, time also loses its meaning, because time is a parameter used to describe the motion and change of an object. Only in a space where there are some bodies moving or changing are space and time closely linked together to form a unified four-dimensional space-time.

We then discuss the second problem of the expansion of the universe: How to measure the expansion speed of the universe? According to the principle of cosmology, there is no center in the universe, and there is a universal cosmic time, so any point can be used to measure the expansion speed of the universe. According to the perspective of inertial frame of center of mass, there is no universal cosmic time. Only when observing from the center of the universe can we correctly describe its expansion, just as only when observing from the sun can we correctly describe the motion of the planets. We have assumed that the earth is close to the center of the universe, so we can easily measure the expansion rate of the universe. Due to the limitation of the speed of light, the speed of a body we observe is not its current speed, but the speed at a certain time in the past. If we observe that the velocity of a particle at position $r$ is $v$, we may think that its velocity before time $t = r / c$ is $v$, since the time required for light to travel over the distance is $t$.

At the beginning of the big bang, all the particles expanded outward at the speed close to the speed of light. Under the action of gravity, the speeds of particles gradually slowed down. The universe has been experiencing decelerated expansion since its creation. Hubble's law is just the demonstration of the decelerated expansion of the universe. As the observational velocities actually represent the velocities of the celestial bodies at past times, it is reasonable to deduce that the farther the distances of the celestial bodies, the faster their receding velocities with respect to the center of the universe, which agrees qualitatively with Hubble's law. However, Hubble's law holds only for the expansion of the universe within a certain range, and is not applicable to the whole universe. In other words, Hubble constant is not a constant, it may vary with different evolution stage of the universe.

The observational data of supernovae showed that the luminosity distances of distant supernovae are larger than their Hubble distances (see, e.g. [23, 24]), which can be explained by the varying Hubble constant, or that the expansion rate of the universe in the past is larger than the current one. So the supernovas evidence has nothing to do with cosmological constant and dark energy. According to the principle of cosmology, the observational data of supernovae indicated the accelerated expansion of the universe. The cosmological constant $\lambda$ in Einstein's gravitational field equation $R_{\mu\nu} - \frac{1}{2} R \cdot g_{\mu\nu} + \lambda \cdot g_{\mu\nu} = -8\pi T_{\mu\nu}$ is usually used to represent the negative pressure field acting as a repulsive force, which is equivalent to dark energy. Dark energy has a strange property. In terms of special relativity, energy is equivalent to mass, so the greater the energy, the larger the gravity, so the uniform conventional energy cannot generate repulsion. The origin of dark energy is often due to vacuum energy in quantum field theory. However, according to the present theory, the vacuum energy is infinite. Even if the vacuum energy is cut off at the Planck energy scale, the vacuum energy density is 120 orders of magnitude larger than the cosmic term density in order to make the theory consistent with observations.

It can be seen that the current dilemma of cosmology comes from the generalization of the Copernicus principle. Based on this principle, there are two generalizations. One is that the center of the universe is the only inertial frame according to the hierarchy of earth → sun → galaxy → galaxy cluster → universe, and the other leads to a universe without a center. The current cosmology chooses the latter one, which is the deep reason for the accelerated expansion of the universe and the introduction of dark energy.

## 9 Conclusion

In general, our theory is equivalent to general relativity in weak gravitational field. Compared with general relativity, our theory is much simpler and easier. In addition, our metric expression does not contain gravitational radius and thus black hole and singularity do not exist in our theory.

Our theory is founded on the inertial frame. The apsidal motion and gravitational radiation of binary stars, and the expansion of the universe, should be observed from their respective frames of center of mass. When observed from the center of mass of the system, the results of the apsidal motion and gravitational radiation in our theory disagree with those of general relativity, and dark energy is not needed.

## Appendix A The source program to compute the mass of neutron star

```
#include "stdio.h"
#include "math.h"

double pi=3.14159265;            //define constant π
double c=2.99792458E8;           //speed of light
double h=6.62606876E-34;         //Planck constant
double G=6.673E-11;              //gravitational constant
double m=1.6749286E-27;          //neutron mass
double m_sun=1.9891E30;          //solar mass
double k=pi*pow(m,4)*pow(c,5)/(4*pow(h,3));    //the value of κ in Eq. (90)
double r;          //the radius of the neutron star
double rou;        //the density of the neutron star
double P;          //the pressure of the neutron star
double M;          //the mass of the neutron star
double t;          //the parameter of equation of state
double step=0.5;   // step length
double f1(double, double, double);   //the expression of differential equation_1
double f2(double, double, double);   // the expression of differential equation_2
double k1[3], k2[3], k3[3], k4[3];      //intermediate variables of Runge-Kutta method
double (*p[2])(double, double, double)={f1,f2};   // Define an array of function pointers so that we
          //can call the corresponding function by index

double ch(double t)
{
    return   (exp(t)+exp(-t))/2;
}

double sh(double t)
{
    return   (exp(t)-exp(-t))/2;
}

double f1(double r, double M, double t)   //the expression of dM/dr
{
```

```
	double val;
	val = 4*pi*r*r*rou;
	return val;
}

double f2(double r, double M, double t)    //the expression of dt/dr
{
	double val1, val2;
	val1 = -3*G*M*(rou+P/(c*c))/(r*r)/(k*(ch(t)-4*ch(t/2)+3));
	val2 = -G/(r*r)*(rou+P/c/c)*(M+4*pi*r*r*r*P/c/c)/(1-2*G*M/(r*c*c))
	       *3/(k*(ch(t)-4*ch(t/2)+3));
	return val1;    // If val2 is returned, the result of TOV equation will be calculated
}

void calc()      // fourth order Runge-Kutta algorithm
{
	for(int i=0; i<2; i++)
	{
	   k1[i+1] = (*p[i])(r, M, t);
	   k2[i+1] = (*p[i])(r+step/2, M+step*k1[1]/2, t+step*k1[2]/2);
	   k3[i+1] = (*p[i])(r+step/2, M+step*k2[1]/2, t+step*k2[2]/2);
	   k4[i+1] = (*p[i])(r+step, M+step*k3[1], t+step*k3[2]);
	}

	M = M+step/6*(k1[1]+2*k2[1]+2*k3[1]+k4[1]);    //calculate the mass of next time
	t = t+step/6*(k1[2]+2*k2[2]+2*k3[2]+k4[2]);       //calculate the value of t of next time
	r = r+step;                                       //calculate radius of next time
	rou = k*(sh(t)-t)/(c*c);
	P = k*(sh(t)-8*sh(t/2)+3*t)/3;
}

void main()
{
	FILE *fp;
	fp = fopen("result.txt","a");
	r = 0.001;    // the initial value of r cannot take 0, otherwise singularity will occur
```

```
    M = 0;        //the initial mass of neutron star is taken as 0
    t = 1;        //different values of t will lead to different masses and radii of neutron star

    while (t>0.001)
       calc();

    fprintf(fp,"%f\t%f", M/m_sun, r/1000);    // Save the results to the result.txt file
    fclose(fp);
}
```